\documentclass[twocolumn,preprintnumbers,amsmath,amssymb,aps,floatfix,superscriptaddress]{revtex4}
\usepackage{epsfig}
\usepackage{mciteplus}
\usepackage{color}
\begin{document}

\title{Statistical mechanics far from equilibrium: prediction and test for a sheared system}

\author{R. M. L. Evans}
\affiliation{School of Physics and Astronomy, University of Leeds, LS2 9JT, U.K.}

\author{R. A. Simha}
\affiliation{Department of Physics, IIT Madras, Chennai 600 036, India.}

\author{A. Baule}
\affiliation{The Rockefeller University, 1230 York Avenue, New York, NY 10065, U.S.A.}

\author{P. D. Olmsted}
\affiliation{School of Physics and Astronomy, University of Leeds, LS2 9JT, U.K.}

\date{28 August 2009}

\begin{abstract}
We report the complete statistical treatment of a system of particles interacting via Newtonian forces in continuous boundary-driven flow, far from equilibrium. By numerically time-stepping the force-balance equations of a model fluid we measure occupancies and transition rates in simulation. The high-shear-rate simulation data verify the invariant quantities predicted by our statistical theory, thus demonstrating that a class of non-equilibrium steady states of matter, namely sheared complex fluids, is amenable to statistical treatment from first principles.
\end{abstract}

\maketitle

Complex fluids relax slowly so their structure is radically re-ordered by flow, as in shear-aligning liquid-crystals \cite{archer1995mtf}, jamming suspensions \cite{haw2004jtf}, or liposome creation \cite{diat93a,*diat93}. 
Sheared fluids consist of particles following the same Newtonian equations of motion as at equilibrium, since no field is applied to drive them; only the boundary conditions differ. 
Nevertheless, they violate equilibrium statistical mechanics \cite{Landau}, and only the distributions of entropy and work have been rigorously analysed in such cases \cite{NatureRef,ScienceRef,Jar2008EPJ,*Jar1997PRL,*Cro1999PR,EVACOH1993PRL,*GALCOH1995PRL,*EvaSea2002AP}.
In processing and using complex fluids, a state of flux is the  
rule rather than the exception, e.g. molten plastic flowing into a  
mould, blood flowing within capillaries, or grease lubricating a  
rotating axle. Under continuous shear flow, these systems exhibit  
statistically steady states with intriguing similarities to  
equilibrium phase behaviour. For example, in ``shear-banding'' of 
worm-like micelles \cite{fielding2007cds,*olmsted2008psb,*spenley93}, the  
fluid itself partitions the applied shear into a region of 
low-viscosity \emph{oriented} material at high strain-rate, coexisting  
with a slower, more viscous region. The parameters controlling this  
structural phase transition are shear rate and concentration, in  
addition to temperature. Typically, simplified models with artificial
dynamics \cite{ArmandBandingFlow,*Santos86} or near-equilibrium approximations 
\cite{Taniguchi04,*CasasV93,*Kawasaki73} are employed in modelling  
these types of system, without knowledge of 
any fundamental principles. Here we validate numerically a complete theory of the detailed statisitics of transition rates and occupancies in a realistic driven system with Newtonian interactions.

We consider a macroscopic region of
fluid, our \emph{system}, embedded in a larger volume of the same
fluid, which acts as a \emph{heat bath} or
\emph{reservoir}, and exerts time-dependent random forces on the system's
boundary that are not predictable from a knowledge of the state of the
system alone. An instantaneous
\emph{microstate} of the system is defined by the exact positions and
momenta of all its constituent particles. The laws of motion governing
its dynamics can be fully summarized by a set $\{\omega_{ab}\}$ of
${\cal N}$ transition rates between every possible pair of microstates
$a$ and $b$ that the system can adopt. Here, $\omega_{ab}$ is the
probability per unit time that the system, currently in microstate
$a$, will be found in microstate $b$ an instant later (so
$\omega_{ab}=0$ for transitions that would violate the laws of
motion). In the presence of random impulses from the reservoir, the
latter microstate is not uniquely determined. Thus, the set
$\{\omega_{ab}\}$ describes \emph{both} the system's dynamics
\emph{and} the probability distribution of forces from the reservoir.

Although forces from the reservoir are stochastic, their randomness is
nonetheless governed by strict rules. The conditions of thermodynamic
\emph{equilibrium} constrain the transition
rates to obey the principle of detailed balance (DB), which states
that the ratio of forward to reverse transition rates between any pair
of microstates must equal the Boltzmann factor of their energy
difference, $\omega^{\rm eq}_{ab}/\omega^{\rm eq}_{ba}=\exp(E_a-E_b)$
(with microstate energies $E_i$ measured in units of the thermal
energy $k_BT$).  In other words, the statistical properties of an equilibrium
reservoir impose ${\cal N}/2$ constraints on the ${\cal N}$ rates (one
per pair). Equivalently, in a Langevin description of the equilibrium
dynamics, the added noise must obey a Fluctuation Dissipation
Theorem (FDT) \cite{gardiner} and is Gaussian, with strength
determined by the thermodynamic temperature. In a sheared steady
state, the system again receives stochastic forces from the reservoir,
but with some non-equilibrium distribution; indeed, those forces make
the system flow. This generates a different set
$\{\omega_{ab}\}$ of transition rates, or equivalently a
different noise distribution.

In the absence of a rigorous theory of non-equilibrium statistical 
mechanics (notwithstanding non-equilibrium generalizations of 
{\em thermodynamics} \cite{OonPan1998PTPS,*SasTas2006JSP}), it has 
become common practice either to invent the non-equilibrium transition 
rates, or equivalently to assume that noise obeys the FDT or some other 
ad hoc criterion (such as colored, \textit{i.e.}  time-correlated noise).
Although the concept of a non-equilibrium temperature is appealing
and has been useful in interpreting non-equilibrium simulations
\cite{haxton2007ada}, statistical derivations typically make
assumpations about microscopic noise or rates
\cite{PhysRevE.55.3898}.  Since arbitrary invention of the
rates is forbidden \emph{at equilibrium} by the constraints of DB,
one might expect the same degree of constraint to arise also from the
statistics of the {\em sheared} reservoir that influences the {\em
non-equilibrium} rates.  Indeed, a non-equilibrium counterpart to DB
can be derived from either information-theoretic
\cite{evans2004rtr,evans2005dbh} or Gibbsian \cite{simha2008pnh}
arguments, which yield a one-to-one mapping between the set of rates
$\{\omega^{\rm eq}_{ab}\}$ in the presence of an equilibrium
reservoir, and those $\{\omega_{ab}\}$ for the same system (with the
same Hamiltonian) bounded by the sheared steady-state reservoir. The
existence of such a mapping means that the driven steady state has the
same number of constraints as the equilibrium case, since one may
first define an equilibrium system (respecting DB), then apply the
mapping to \emph{derive} the driven dynamics.

For completeness, let us briefly discuss the ingredients required to
derive the mapping \cite{evans2005dbh,simha2008pnh}. The transitions 
are nontrivially correlated via the dynamics so, instead of microstate
\emph{transitions}, the basic objects for statistical analysis are
\emph{phase-space paths} \cite{PhysRevE.61.2361,*komatsu2008esd}. Such a path
describes a system's entire history of microstates; the
position and momentum of every constituent particle at every instant
during the very long duration of some steady-state experiment.  A
phase-space path $\Gamma$ is therefore a complete description of the
real physics exhibited by the system for a particular realisation of
the noise, including all existing non-trivial spatial and temporal
correlations.

We conceptually construct an \emph{ensemble} of weakly interacting
systems that can (i)~exchange energy (as in the equilibrium canonical
ensemble) and (ii)~arbitrarily distribute the total shear strain
amongst the member systems (as occurs amongst a set of finite fluid
elements within a larger volume of fluid undergoing shear-banding).
The crucial property of the phase-space paths of this ensemble is that
they are uncorrelated with each other, except via those two quantities 
(energy and shear strain) that are exchanged over long range due to 
local conservation laws. Despite this lack of correlation, each path 
fully describes all the spatial and temporal
correlations of its system. We consider all paths consistent
with the laws of motion, but differing in their realisations of the 
noise. For any such set of {\em uncorrelated}
objects, their exact probability distribution $p_\Gamma$ can be 
found \cite{mandl} by maximizing the entropy of that
distribution (not to be confused with a thermodynamic entropy) subject
to constraints from normalization and conservation laws. The
derivation for the sheared case differs from that for equilibrium by
only one additional constraint, fixing the ensemble-average of the
total shear strain attained by the systems over the duration of the
thought-experiment.  The result \cite{simha2008pnh} is a relationship
between the path distributions in the two ensembles,
\mbox{$p_\Gamma\propto p_\Gamma^{\rm eq}\exp \nu\gamma_\Gamma$}, where
$\nu$ is a Lagrange multiplier for the extra shear constraint, and
$\gamma_\Gamma$ is the total strain for path $\Gamma$. By summing this
relationship over all paths that contain a given transition between a
pair of microstates, one obtains the exact mapping, discussed above,
between the sets of transition rates $\{\omega_{ab}\}$ and
$\{\omega^{\rm eq}_{ab}\}$. (Supplementary material
demonstrates how the same analysis can be used to obtain microstate
occupancies, a non-equilibrium analogue of Boltzmann's law.)

We have not stated the one-to-one mapping explicitly here, as it would
require further notation to be introduced. However, it was recently
noticed \cite{bauleevans2008,evans2005dbh} that the mapping
implies some remarkably simple relationships that, being exact,
apply arbitrarily far from equilibrium. These relationships apply
to a state-space with arbitrary connectivity between any set
of microstates:
\begin{enumerate}
\item The total exit rate from any given microstate differs from its
  equilibrium value by a shear-rate-dependent constant that is the
  same for all microstates, i.e. \mbox{$\sum_b \left(
      \omega_{ab}-\omega^{\rm eq}_{ab} \right)=Q(\nu)\;\forall\;a$}.
\item The product of forward and reverse transition rates is the same
  in the equilibrium and sheared ensembles, i.e.
  \mbox{$\omega_{ab}\,\omega_{ba}=\omega^{\rm eq}_{ab}\,\omega^{\rm
      eq}_{ba}\;\forall\;a,b$}.
\end{enumerate}

\begin{figure}
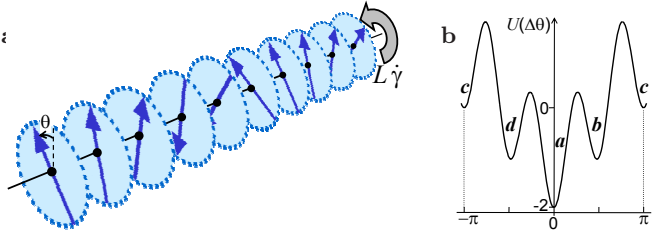

  \begin{center}
      {\bf \raisebox{2.5cm}{a$\!\!\!\!$}} \epsfxsize=5.3cm\epsffile{rotorfigpadded.eps}
      {\bf \raisebox{2.5cm}{\quad b$\!\!$}} \epsfxsize=2.6cm\epsffile{Ufig.eps}
    \caption{\label{rotorfig}\small (a) The one-dimensional rotor
      model of length $L$. Each of the $L$ rotors is characterized by
      its angle $\theta$ and angular velocity $\dot{\theta}$. (b) The
      potential of interaction between neighbours,
      $U(\Delta\theta)=-\cos\Delta\theta-\cos 4\Delta\theta$, a
      symmetric periodic function of their angular difference
      $\Delta\theta$, with four wells, labelled
      $a,b,c,d$.\vspace{-4mm}}
\end{center}
\end{figure}

We have tested the above theory in a one-dimensional model system, a
``fluid'' of rotors (see Fig.~\ref{rotorfig}a), each interacting with
its neighbours via torsional forces, and respecting Newton's laws of
motion. The angular acceleration of each rotor is proportional to its net
unbalanced torque, which is the difference between the torques applied
by its two neighbours. The torque between neighbours has three
contributions: conservative, dissipative and random. The conservative
part is the gradient of the four-well potential $U(\Delta\theta)$
shown in Fig.~\ref{rotorfig}b, which is a function of the angular
difference between the rotors. Thus, the zero-temperature ground-state
has all rotors parallel, but antiparallel and perpendicular
configurations are also moderately favourable.  The uncorrelated
random contribution to the torque 
(representing any microscopic degrees of freedom that are independent 
of shear strain, {\em e.g.}~Brownian forces from solvent molecules) 
has a uniform distribution with
width $\sigma$ and zero mean. The dissipative part is proportional to
the difference in angular velocity between neighbours, with a constant
of proportionality that can be rescaled to unity without loss of
generality, leaving the noise strength $\sigma$ as the model's only
parameter.  Neighbours experience equal and opposite torques, so that
angular momentum is exactly conserved in the model. The boundary
conditions are periodic, and the equations of motion are numerically
time-stepped.

As described thus far, this is an equilibrium model. Once initial
transients have died away, it exhibits Boltzmann statistics in the
occupancies of the potential $U(\Delta\theta)$. At sufficiently low
noise strength, the relative angles between neighbours are mostly
confined close to the local potential minima, with only occasional
transitions between potential wells. The {\em measured} rates of those
transitions respect DB. Note that DB is not imposed \emph{a priori};
it emerges from the dynamics at equilibrium.

We model a fluid under (angular) shear by \emph{twisting} the model
(see Fig.~\ref{rotorfig}a). The twist is imposed via the periodic
boundary condition, by introducing an offset in the angle measured
between rotors either side of the boundary, and increasing that offset
linearly in time, at a rate $L\dot{\gamma}$, so that the twist rate
per rotor is $\dot{\gamma}$. As with ordinary periodic boundary
conditions, this angular analogue of Lees-Edwards boundaries
\cite{leesedwards} avoids edge effects, since it is a non-local
condition on the topology of the space; the rotors cannot tell where
the boundary is located, as they are only aware of \emph{relative}
angles.

To apply our general theory, some region of the model must be
defined as the \emph{system}, while the large remainder is the
\emph{reservoir}, supplying unpredictable non-equilibrium forces to
it. The system should be much larger than any correlation length, to
ensure \emph{weak} coupling to an \emph{uncorrelated} reservoir.
Unfortunately, a large system implies a high-dimensional phase space
($\theta$ and $\dot{\theta}$ for each rotor), so that acquiring a
statistically significant sample of all the transition frequencies
becomes prohibitively time-consuming. We take two steps to
reduce the phase space.  First, we take the limit of small moment of
inertia, so that momenta are no longer independent, and the phase
space reduces to the set of inter-rotor angles $\Delta\theta$. This
has the added advantage, in a one-dimensional force-chain, of reducing
the correlation length to zero since, with vanishing rate of change of 
angular momentum, the forces now balance globally. We are therefore 
able, secondly, to treat every
inter-rotor gap (with its single characteristic variable
$\Delta\theta$) as a system, each surrounded by a non-equilibrium
reservoir. Note that the general theory applies to systems with
non-vanishing correlation lengths, and we are treating a special case
only for the sake of expediency. An impediment still exists for
analysing this rotor model: the general theory applies to transitions
between microstates, whereas the measured transition rates are between
potential wells $a$, $b$, $c$ and $d$, that are finite in extent. \emph{If}
these four \emph{continuous sets} of microstates are sufficiently
analogous to true microstates, then the theory applies in this case.
Subject to that qualification, we can use the model to test the
theory's central assumption of \emph{ergodicity};
\textit{i.e.}~the available phase-space paths are representatively
sampled by the dynamics.

\begin{figure}
  \epsfxsize=8.8cm\epsffile{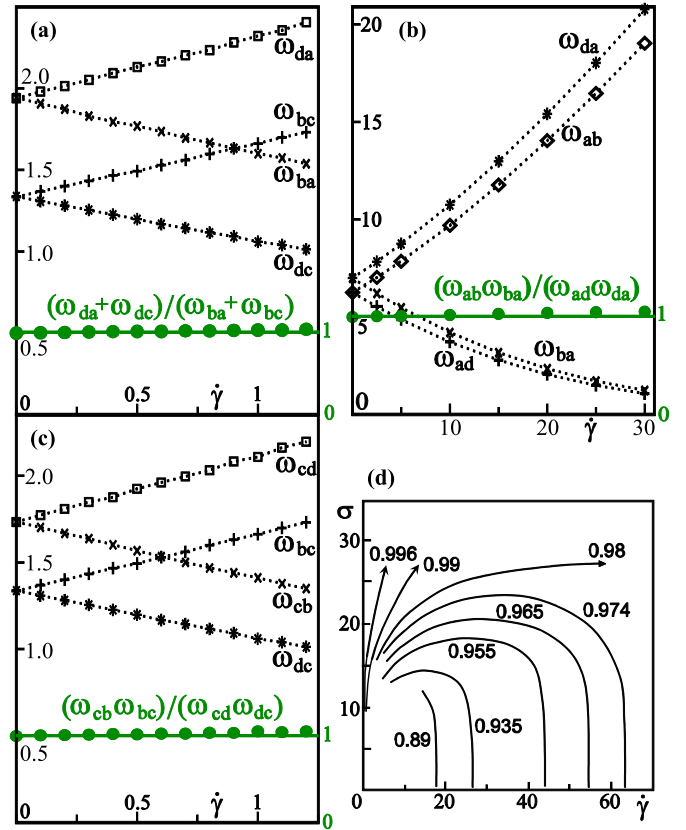}
     \caption{\label{sumfig}\small (a) Test of the prediction
      $\omega_{da}+\omega_{dc}=\omega_{ba}+\omega_{bc}\;\forall\;\dot{\gamma}$
      with noise strength $\sigma=10$. The left-hand ordinate measures rates,
      using the same units as the abscissa, while the right-hand ordinate 
      measures the dimensionless ratio. (b) Test of the predicted relationship
      $\omega_{ab}\,\omega_{ba}=\omega_{ad}\,\omega_{da}\;\forall\;\dot{\gamma}$,
      with $\sigma=20$. At this higher noise strength, higher shear rates are 
      numerically accessible because of the greater number of observed backward
      transitions. (c) Test of the prediction
      $\omega_{cd}\,\omega_{dc}=\omega_{cb}\,\omega_{bc}\;\forall\;\dot{\gamma}$,
      for the same parameters as in (a). (d) Contours of the measured ratio
      $(\omega_{ba}+\omega_{bc})/(\omega_{da}+\omega_{dc})$, predicted
      to be unity across the whole parameter space of shear rate
      $\dot{\gamma}$ and noise strength $\sigma$. Data acquisition
      time limited simulations in the bottom left-hand corner. See supplementary 
      material for more data}
\end{figure}

We first test the predicted relationship between exit rates.  The
required quantity $\sum_a \omega^{\rm eq}_{ab}$ implicitly depends on
the unknown temperature of the compared equilibrium system. However,
we can eliminate that unknown by appealing to a symmetry of
$U(\Delta\theta)$.  Since wells $b$ and $d$ are identical \emph{at
equilibrium}, they have equal total exit rates, 
\mbox{$\omega^{\rm eq}_{ba}+\omega^{\rm eq}_{bc}=\omega^{\rm eq}_{da}
+\omega^{\rm eq}_{dc}$}. 
So relationship number 1 predicts them also to have
equal total exit rates in the driven case:
\mbox{$\omega_{ba}+\omega_{bc}=\omega_{da}+\omega_{dc}$} \emph{for all} 
imposed shear rates $\dot{\gamma}$.  
This equality is not 
obvious, since the equilibrium symmetry is broken in the
driven case, where one of the potential wells is upstream of the
other. Measurements of the four rates in question are plotted against
shear rate in Fig.~\ref{sumfig}a for a particular noise strength.
The rates vary considerably with $\dot{\gamma}$ and depart
significantly from their equilibrium values. Nevertheless, the ratio
of sums, as anticipated, remains very close to unity.

Next we test the second predicted relationship, between products of
rates. Again, we exploit the symmetries of the hypothetical
equilibrium state to obtain a relationship between the measured rates
in the actual driven system only. In \emph{equilibrium},
symmetry of $U(\Delta\theta)$ (together with DB) implies $\omega^{\rm
  eq}_{ab}=\omega^{\rm eq}_{ad}$ and $\omega^{\rm eq}_{ba}=\omega^{\rm
  eq}_{da}$. Substitution into proposed relationship 2 implies a
constraint on the measured rates in the driven system:
$\omega_{ab}\,\omega_{ba}=\omega_{ad}\,\omega_{da}\;\forall\;\dot{\gamma}$.
This prediction is tested in Fig.~\ref{sumfig}b: as before, while
the individual rates vary significantly across the range of driving
speeds, the prediction is obeyed to an excellent approximation.
Similarly, the equilibrium symmetry about well $c$ gives rise to a
third non-equilibrium prediction,
$\omega_{cd}\,\omega_{dc}=\omega_{cb}\,\omega_{bc}\;\forall\;\dot{\gamma}$,
verified in Fig.~\ref{sumfig}c.

To quantify the accuracy of the non-equilibrium theory across the
model's whole parameter space, contours of the measured ratio
$(\omega_{ba}+\omega_{bc})/(\omega_{da}+\omega_{dc})$ are plotted in
Fig.~\ref{sumfig}d. The theory predicts a value of
unity everywhere. The small but significant discrepancies between
theory and data may be due to the finite extents of the states
$a,b,c,d$, making them not true microstates. Although the
non-equilibrium rotor model constitutes an imperfect test of the
theory, the theory performs strikingly well here, and exceeds the
predictive power of any approximate methods available to
non-equilibrium statistical mechanics. Importantly, the discrepancies
do not increase with $\dot{\gamma}$, as would be the case for a
near-equilibrium theory. The discrepancies at low $\sigma$ may
alternatively be due to failure of the ergodic hypothesis. The theory
shares this hypothesis with equilibrium statistical mechanics, which
also fails for non-ergodic systems such as glasses, but is successful
for a wide range of applications.

While the logistics of data acquisition has restricted our study of
the rotor model to the zero-mass, zero-correlation-length limit, the
theory should also apply to the more general case with momentum
degrees of freedom, thus encompassing phases with non-zero correlation
lengths. However, even the testable model studied here exhibits highly
non-trivial behaviour, excellently predicted by the non-equilibrium
statistical mechanical theory. Of course, no theory can describe {\em all}
non-equilibrium steady states, since such states are much more diverse than the set of all equilibrium problems, for instance encompassing molecular motors, convection cells, granular media and traffic flow. 
We have nevertheless presented a fundamental theory that governs the steady-state motion of any flowing system on which work
is done by a weakly-coupled non-equilibrium reservoir that is ergodic
(sufficiently mobile to explore its space of available states
thoroughly), and microscopically reversible (containing normal
particles with no sense of direction). Such systems include all
sheared complex fluids, whose phenomenology is as important to
future technologies as it is to our understanding of non-equilibrium
physics.

\noindent{\bf Acknowledgements:} The work was funded by EPSRC grant GR/T24593/01. RMLE is supported by the Royal Society.


\end{document}